# In-Band Co-Polarization Scattering Beam Scanning of Antenna Array Based on 1-Bit Reconfigurable Load Impedance

Binchao Zhang, *Member, IEEE*, Fan Yang, *Fellow, IEEE*, Shenheng Xu, *Member, IEEE*, Maokun Li, *Senior Member, IEEE,* and Cheng Jin, *Senior Member, IEEE*

*Abstract*- Controlling the in-band co-polarization scattering of the antenna while maintaining its radiation performance is crucial for the low observable platform. Thus, this paper studies the in-band co-polarization scattering beam scanning of antenna arrays. Firstly, the regulation method of antenna scattering is analyzed theoretically, concluding that the amplitude and phase of the antenna's scattering field can be regulated by changing the load impedance. Subsequently, PIN diodes are implemented to control the load impedance of the antenna. Consequently, the scattering of the antenna, ensuring that the antenna's scattering meets the condition of equal amplitude and a phase difference of $180^o$ when the PIN diode switches, thereby realizing scattering beam scanning. Moreover, by introducing an additional pre-phase, the inherent symmetric dual-beam issue observed in traditional 1-bit reconfigurable structures is overcome, achieving single-beam scanning of the scattering. Finally, a 1×16 linear antenna array is designed and fabricated, which operates at 6 GHz with radiation gain of 16.3 dBi. The scattering beams of the designed array can point to arbitrary angles within $\pm 45^o$, significantly reducing the in-band co-polarization backward radar cross section. The measured results align well with the simulated ones.

## I INTRODUCTION

With the advancement of electromagnetic stealth technology, the radar cross section (RCS) of equipment platforms has been substantially reduced, leading various types of antenna arrays mounted on the platform to become the primary contributors to the RCS. These platforms often face radar detection from either allies or adversaries [1]. Antenna arrays with reconfigurable RCS can dynamically enhance the RCS for ally detection or reduce it for adversary detection by reshaping the scattering pattern. Hence, developing dynamically reconfigurable in-band co-polarization scattering fields, while ensuring antenna radiation performance, is of paramount importance. However, due to radiation constraints, reconstructing the in-band co-polarization scattering field remains a substantial challenge.

Traditionally, research predominantly concentrated on reducing the out-of-band RCS of antennas, and aiming to enhance performance concerning large angle [2], wide bandwidth [3], and low profile [4], among others. Common approaches involved employing absorbers [5]-[7], artificial magnetic conductors [8]-[10], frequency selective reflection/transmission structures [11]-[15], to effectively reduce the out-of-band RCS while preserving the antenna's radiation performance.

However, advancements in radar detection technology necessitate further reduction of in-band RCS of antennas. Achieving in-band RCS reduction is more challenging compared to out-of-band RCS, as the co-polarization radiation and scattering fields of the antenna interfere with each other within the same frequency band. Some studies have successfully managed to reduce in-band RCS without compromising antenna radiation performance [16]-[23]. Common strategies involve implementing additional impedance matching networks or integrating metasurfaces. Specifically, impedance matching networks can regulate the antenna mode scattering field to cancel out the structural mode scattering field, thereby reducing the in-band RCS [24]-[26]. Alternatively, integrating metasurfaces and antennas endows the antenna with the capability to control in-band RCS [27], [28].

Additionally, several studies have been conducted to achieve antennas' reconfigurable scattering fields. One notable study presented a novel Fabry-Pérot antenna that employs a multifunctional metasurface overlay on the antenna aperture to attain high-gain radiation characteristics and in-band scattering control [29]. However, incorporating an additional metasurface structure elevates the complexity of the antenna profile. Another innovative design introduced a new patch antenna array featuring a reconfigurable loop structure [30], achieving high radiation efficiency along with a reconfigurable scattering field. The in-band RCS of the antenna can be considerably reduced by adjusting the null position of the scattering pattern. The 1-bit reconfiguration of in-band scattering field was explored in [31], where adjusting the input impedance of the antenna allowed for control of the scattering field. Nevertheless, the inherent symmetric dual-beam issue of the 1-bit reconfigurable structure prevented the attainment of single-beam control of the in-band scattering field. Hence, more in-depth research is imperative to accomplish single-beam control of co-polarization and in-band scattering, while preserving the antenna radiation performance

To verify the method, a 1×16 linear antenna array is designed, fabricated and measured, which shows that the proposed method successfully achieves single-beam scanning of in-band co-polarization scattering and stable radiation.

This work is supported in part by the National Key Research and Development Program of China under Grant 2020YFB1806302, in part by THE XPLORER PRIZE, and in part by China Postdoctoral Science Foundation under Grant 2023M731886. (*Corresponding author: Fan Yang*)

B. Zhang, F. Yang, S. Xu and M. Li are with the Department of Electronic Engineering, Tsinghua University, Beijing 100084, China. (Email: zhangbinchao@tsinghua.edu.cn, fan yang@tsinghua.edu.cn)

C. Jin is with the school of Cyberspace Science and Technology, Beijing Institute of Technology, Beijing 100081, China.

Therefore, the 1-bit reconfigurable load impedance is employed to manipulate the in-band co-polarization scattering of the antenna array, achieving single beam scanning. PIN diodes are embedded to ensure the scattering meets the condition of equal amplitude and a $180^o$ phase difference, thus facilitating 1-bit reconfigurable scattering beam control. Moreover, an additional pre-phase is introduced to resolve the inherent symmetric dual-beam issue found in traditional 1-bit reconfiguration structures, enabling the attainment of single-beam scanning performance. To validate this method, a 1×16 linear antenna array is designed, fabricated, and measured. The results demonstrate that the proposed method successfully accomplishes single-beam scanning of in-band co-polarization scattering along with stable radiation performance.

## II THEORETICAL ANALYSIS

To effectively control the in-band co-polarization scattering field, analyzing the relationship between the antenna's radiation and scattering is essential. It is well known that the scattering field $\vec{E}^s$ of an antenna typically comprises two components: the structural mode scattering field $\vec{E}^{ss}$, primarily dependent on the antenna's structure and independent of the antenna's load impedance; and the antenna mode scattering field $\vec{E}^{as}$, originating from power reflections due to the mismatch between load impedance and antenna impedance, subsequently re-radiated by the antenna [32]. These two scattering fields can be related to antenna impedance $Z_A$ and load impedance $Z_L$ through the subsequent equation

$$\vec{E}^s = \vec{E}^{as} + \vec{E}^{ss}, \quad (1)$$

$$\vec{E}^{as}(Z_L) = \frac{\Gamma_L}{1 - \Gamma_a \Gamma_L} \frac{1 - \Gamma_a^2}{2} [\vec{E}^s(\infty) - \vec{E}^s(0)], \quad (2)$$

$$\vec{E}^{ss}(Z_c) = \frac{(1 - \Gamma_a)\vec{E}^s(\infty) + (1 + \Gamma_a)\vec{E}^s(0)}{2}. \quad (3)$$

Here, $\vec{E}^s(\infty)$ and $\vec{E}^s(0)$ denote the scattering fields when the antenna load is open-circuited and short-circuited, respectively. $\Gamma_L$ and $\Gamma_a$ represent the reflection coefficients for the load and the antenna, respectively. They are defined as follows

$$\Gamma_L = \frac{Z_L - Z_c}{Z_L + Z_c}, \quad \Gamma_a = \frac{Z_A - Z_c}{Z_A + Z_c}, \quad (4)$$

Where $Z_c$ represents the characteristic impedance of the transmission line. By substituting equations (2) to (4) into (1), it can be further simplified to

$$\vec{E}^s(Z_L) = \frac{Z_L}{Z_A + Z_L}\vec{E}^s(\infty) + \frac{Z_A}{Z_A + Z_L}\vec{E}^s(0). \quad (5)$$

From equation (5), it can be deduced that the amplitude and phase of the antenna's in-band co-polarization scattering field are exclusively dependent on the load impedance $Z_L$ and the antenna impedance $Z_A$. Given a fixed antenna structur, $Z_A$, $\vec{E}^s(\infty)$ and $\vec{E}^s(0)$ are all constants. Consequently, by varying the load impedance $Z_L$, the amplitude and phase of the antenna's scattering field can be controlled. This enables the realization of the 1-bit reconfigurable feature for the antenna's in-band co-polarization scattering field, thereby facilitating dynamic beam scanning of the scattering.

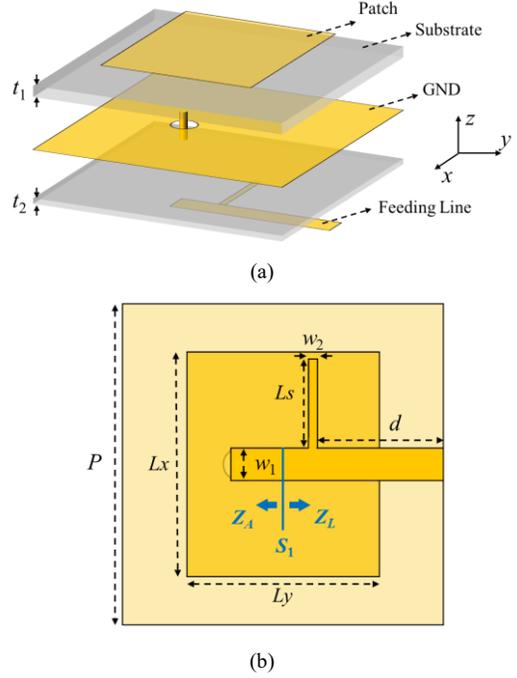

Fig. 1 Illustration of the designed antenna element, (a) perspective view and (b) bottom view.

## III DUAL-BEAM SCANNING OF SCATTERING

In this section, the theoretical analysis is validated using a conventional patch antenna, the structure of which is depicted in Fig. 1. The substrate is made of Rogers RO3003 material with a dielectric constant of 3, and their thicknesses are $t_1 = 2$ mm and $t_2 = 1$ mm, respectively. The operating frequency of the antenna is designed to be $f_0 = 6$ GHz, resulting in patch dimensions of $L_x = 15$ mm and $L_y = 13$ mm, and an element period of half-wavelength, $P = 25$ mm. Additionally, by cascading an open-circuited transmission line of different lengths $L_s$ on the feed line at a distance $d = 10$ mm from the load, the load impedance $Z_L$ can be varied, thus gaining the capability to manipulate the amplitude and phase of the antenna's in-band co-polarization scattering field.

Without loss of generality, a reference plane, denoted as $S_1$, is established at a distance $d_s = 15$ mm from the load. From this reference plane, the antenna impedance $Z_A$ is seen from $S_1$ towards left antenna side, and the load impedance $Z_L$ is seen from $S_1$ towards right load side. Consequently, the antenna impedance is simulated to be $Z_A = 60.55 + j15.63$, and the scattering fields $\vec{E}^s(\infty)$ and $\vec{E}^s(0)$ can be obtained through full-wave simulation. Furthermore, the relationship between the load impedance $Z_L$ and the cascaded open-circuited transmission line length $L_s$ is

$$Z_L = Z_{C1} \frac{Z'_L + jZ_{C1} \tan \sqrt{\varepsilon_r}\beta(d_s - d)}{Z_{C1} + jZ'_L \tan \sqrt{\varepsilon_r}\beta(d_s - d)}, \quad (6)$$

$$Z'_L = Z_{C1} \frac{Z_0 + jZ_{C1} \tan \sqrt{\varepsilon_r}\beta d}{Z_{C1} + jZ_0 \tan \sqrt{\varepsilon_r}\beta d} // -j\frac{Z_{C2}}{\tan \sqrt{\varepsilon_r}\beta L_s}. \quad (7)$$

Here, $Z_{C1}$ and $Z_{C2}$ represent the characteristic impedance of the transmission lines with widths $w_1 = 2$ mm and $w_2 = 0.5$ mm respectively, $\varepsilon_r$ is the relative dielectric constant of the substrate, $\beta = 2\pi/\lambda_0$ is the propagation constant, and $Z_0$ is 50 Ω. Then, the amplitude and phase of the in-band copolarization scattering field can be calculated using equations (5) to (7).

amplitude and phase of the scattering field. As evident in Fig. 2, there is substantial agreement between calculated and simulated amplitude and phase.

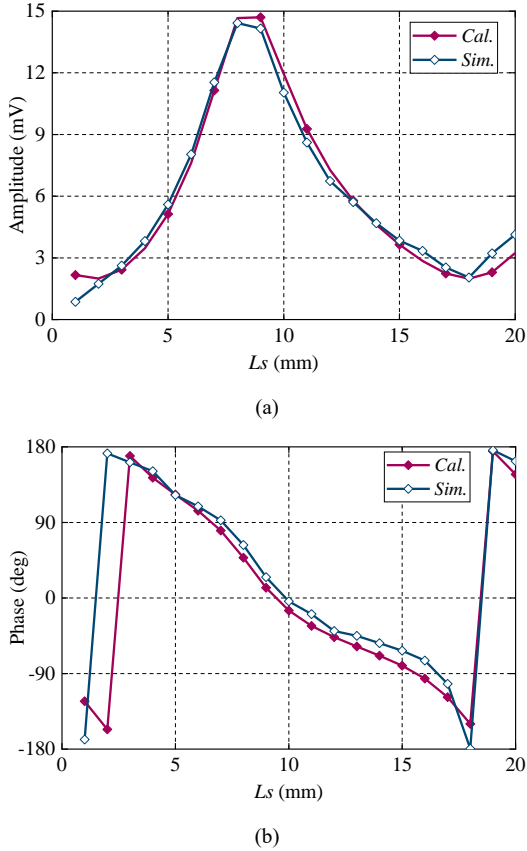

Fig. 2. Comparison of simulated and calculated results of the scattering field as $L_s$ changes, (a) amplitude, (b) phase.

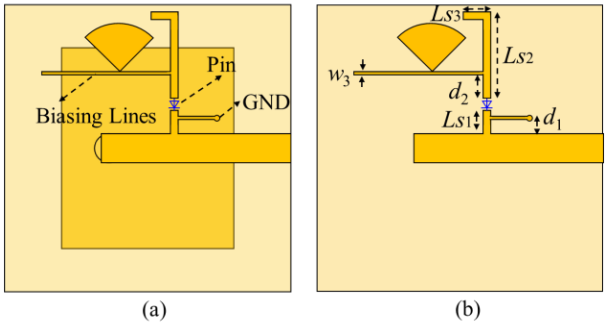

Fig. 3. Designed antenna element with integrated PIN diode, (a) top view and (b) detailed dimensions.

As illustrated in Fig. 2, as $L_s$ varies from 0 mm to 20 mm, the phase of antenna's scattering field can cover $360^o$, with the amplitude initially increasing and subsequently decreasing. Thus, by changing $L_s$ $L_s$, two bit states of the scattering field can be identified, where the amplitudes are equal, and the phases exhibit a $180^o$ discrepancy. This inference is corroborated by juxtaposing simulation outcomes with theoretical calculations. The antenna element depicted in Fig. 1 is subject to periodic boundary conditions, and a normally incident plane wave serves as the excitation to acquire the simulated

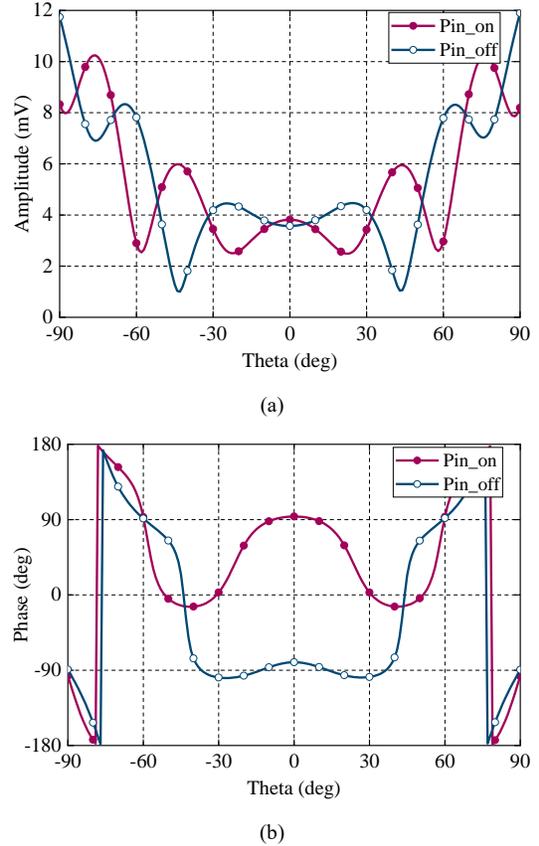

Fig. 4. Simulated amplitude and phase of the scattering field as PIN diode switches, (a) amplitude, (b) phase.

To achieve 1-bit reconfigurable scattering field, a PIN diode is integrated into the cascaded open-circuit transmission line, as illustrated in Fig. 3. The state switch of the PIN diode alters the transmission line length, enabling 1-bit electrical reconfigurability of the scattering field. For simulation, the equivalent circuit model parameters of the PIN diode are: $R_{on} = 1\Omega$, $L_{on} = 0.45$ nH, $R_{off} = 10\,\Omega$, $L_{off} = 0.45$ nH, $C_{off} = 0.126$ pF [33]. The optimized structural dimensions are: $L_{s1} = 2$ mm, $L_{s2} = 8$ mm, $L_{s3} = 2$ mm, $d_1 = 1$ mm, $d_2 = 2.5$ mm, $w_3 = 0.2$ mm. Similarly, the element is simulated in HFSS to obtain the amplitude and phase when the PIN diode switches, as depicted in Fig. 4. The results demonstrate that the amplitude of the scattering field at a $0^o$ elevation angle remains essentially constant when the state of the PIN diode is altered, while the phase experiences a difference of approximately $180^o$, hence substantiating the 1-bit reconfigurable attribute.

Then, a 1×16 antenna array is assembled, utilizing the element depicted in Fig. 3. To calculate the phase distribution needed for scattering beam scanning to $\theta_0$ under the conditions of normal plane wave incidence, the ensuing equation is employed

$$\phi_m = -\frac{2\pi}{\lambda_0} mP \sin\theta_0, m \in (1,16), \qquad (8)$$

Where $\phi_m$ represents the actual compensation phase of each element at different beam scanning angles, $m$ is the serial number of the one-dimensional array element, and $P$ is the period. Subsequently, using equation (9), $\phi_m$ is normalized to the required bit state

$$bit_m = \begin{cases} 0 & \phi_m \in [0,\pi] \pm 2n\pi \\ 1 & \phi_m \in [\pi, 2\pi] \pm 2n\pi \end{cases}. \quad (9)$$

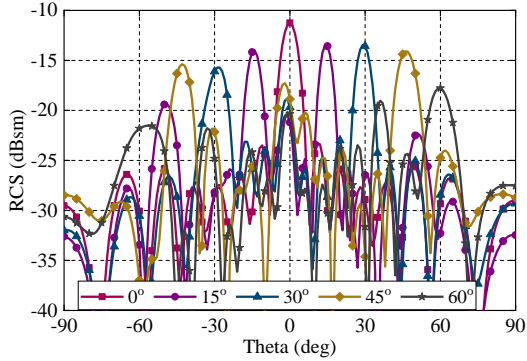

Fig. 5. Simulated scattering dual-beam scanning of the one-dimensional antenna array.

Fig. 5 shows the simulated scattering dual-beam scanning of the one-dimensional antenna array. The depiction reveals that, through varying coding sequences, the in-band co-polarization scattering can be manipulated to not only effectively minimize the backward RCS but also to enhance the RCS at directions of $\pm\theta_0$. Nonetheless, the 1-bit reconfigurable structure will inevitably produce symmetric dual scattering beams under normal incident plane wave, which will result in large RCS in unwanted directions. Hence, further exploration and meticulous design are essential to attain single-beam 1-bit reconfiguration of the in-band co-polarization scattering.

## III SINGLE-BEAM SCANNING OF SCATTERING

According to [34], when a normal plane wave is incident, the normalized array factor of the antenna array is

$$|AF(\theta)| = \frac{1}{N}\sqrt{\sum_{m=1}^{N}\sum_{n=1}^{N}\cos\left(\frac{2\pi(m-n)P}{\lambda_0}\sin\theta + (\phi_m - \phi_n)\right)}, \quad (10)$$

where $N$ is the number of array elements, and both $m$ and $n$ are element indices. In the 1-bit case, $(\phi_m - \phi_n) = 0, \pm\pi$, thus equation (10) can be simplified as

$$|AF(\theta)| = \frac{1}{N}\sqrt{\sum_{m=1}^{N}\sum_{n=1}^{N}(-1)^u\cos\left(\frac{2\pi(m-n)d}{\lambda}\sin\theta\right)}, \quad (11)$$

$$u = \begin{cases} 0, & \phi_m - \phi_n = 0 \\ 1, & \phi_m - \phi_n = \pm\pi \end{cases}. \quad (12)$$

The array factor of the 1-bit reconfigurable antenna array is observable as an even function, yielding two inherent symmetric scattering beams.

In order to achieve the scanning of a single scattering beam, it's imperative to introduce two distinct pre-phases to ensure that $|AF(\theta)|$ is neither odd nor even, subsequently disrupting the symmetrical dual beams. For illustrative clarity, the two different pre-phases can be selected as 0 and $\pi/2$. For the $1\times16$ antenna array, the random pre-phase distribution is

$$\Phi^{pre} = \frac{\pi}{2}\text{randi}([0,1], 1, 16). \quad (13)$$

Then, the 1-bit phase distribution for each element is

$$\phi_m^{pre} = \phi_m + \Phi_m^{pre}, m \in (1, 16), \quad (14)$$

$$bit_m^{pre} = \begin{cases} 0 & \phi_m^{pre} \in [0,\pi] \pm 2n\pi \\ 1 & \phi_m^{pre} \in [\pi, 2\pi] \pm 2n\pi \end{cases}. \quad (15)$$

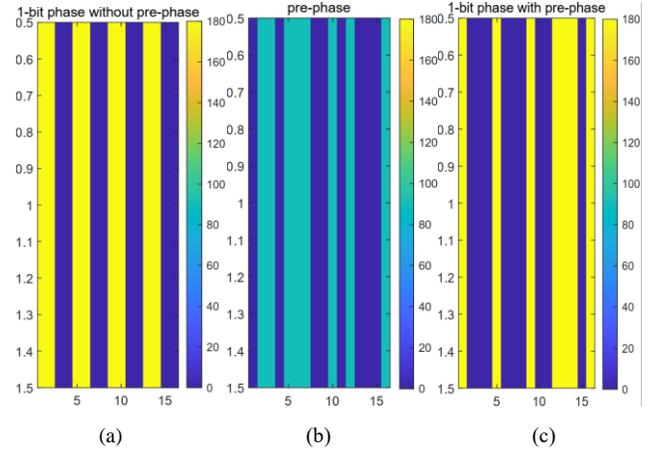

Fig. 6. One-dimensional antenna array with 1-bit phase distribution: (a) without pre-phase, (b) pre-phase, and (c) with pre-phase.

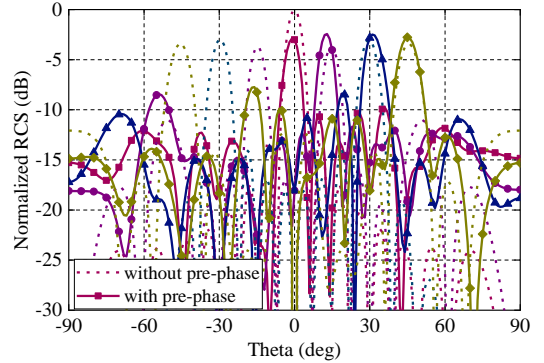

Fig. 7. Comparison of the calculated RCS for symmetric dual-beam and single-beam scanning

Using $\theta_0 = 30^o$ as a representation, Fig. 6 illustrates the 1-bit phase distribution of the one-dimensional array antenna both with and without the pre-phase. It is evident that the incorporation of the pre-phase disrupts the symmetry of the phase distribution, thereby neutralizing the inherent symmetrical dual-beam characteristic. Fig. 7 conveys a theoretical contrast of the dual- and single-beam scattering scanning performance of the $1\times16$ array. It is unmistakably observable that the inclusion of the pre-phase realizes good single-beam scattering scanning performance, and the RCS in the backward direction has a reduction of more than 15 dB.

To achieve the anticipated pre-phase, two distinct element structures, namely I and II, need to be designed. These structures must fulfill the subsequent conditions when the PIN diode switches

$$\begin{cases} Amp_I^{on} = Amp_I^{off} = Amp_{II}^{on} = Amp_{II}^{off} \\ \varphi_I^{on} - \varphi_I^{off} = \varphi_{II}^{on} - \varphi_{II}^{off} = \pi \\ \varphi_I^{on} - \varphi_{II}^{on} = \varphi_I^{off} - \varphi_{II}^{off} = \pi/2 \end{cases}. \quad (16)$$

According to equations (6) and (7), it is evident that the impedance $Z_L$ varies linearly with the distance $d$ of the cascading open-circuit transmission line from the load. Fig. 8 provides the simulated results of the variation of scattering field amplitude and phase with parameter $d$. It is observed that the phase of the scattering field changes linearly with the increase of $d$, and there is always a phase difference of $180^o$ when the PIN diode is on or off. The amplitude of the scattering field undergoes comparatively minor alterations and essentially satisfies the condition of equal amplitude. Hence, element I and II can be constructed by selecting $d = 11$mm and $d = 14.5$ mm, respectively, to ensure that the scattering field amplitude and phase of the two elements meet the conditions given in equation (16).

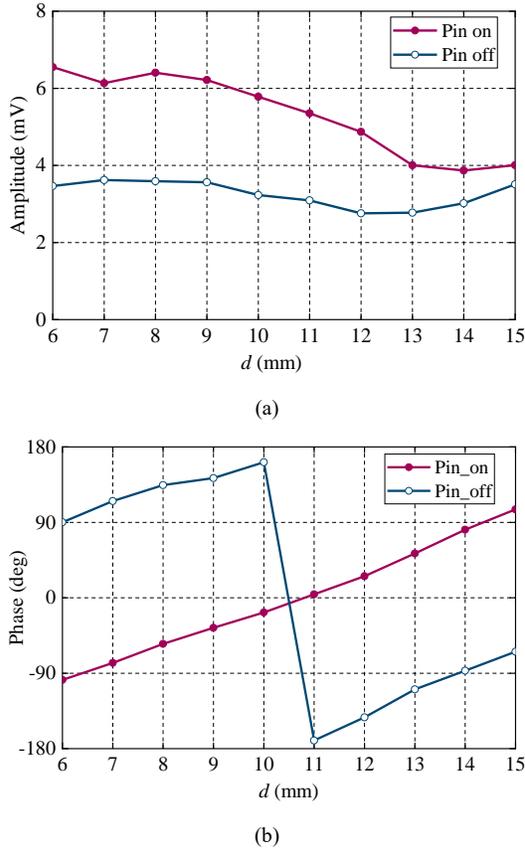

Fig. 8. Variation of scattering field amplitude and phase with parameter $d$, (a) amplitude, (b) phase.

Further, the scattering performance of elements I and II under normal incident plane wave excitation and the radiation performance under discrete port excitation are simulated, as depicted in Fig. 9. In regard to scattering performance, as shown in Figs. 9(a) and 9(b), it is observable that both elements I and II achieve a $180^o$ phase difference when the PIN diode switches, and the scattering phase difference between the two elements is $90^o$. Concurrently, the scattering field amplitudes for all four scenarios are roughly equivalent, thus meeting the quasi-2-bit condition. Pertaining to radiation performance, as illustrated in Figs. 9(c) and 9(d), the reflection coefficients of element I and II are essentially below -10 dB at 6 GHz, regardless of whether the PIN diode is on or off, and the realized gains are around 5 dBi. It indicates that when performing single-beam scanning of the in-band co-polarization scattering, the radiation performance of the antenna remains largely unchanged.

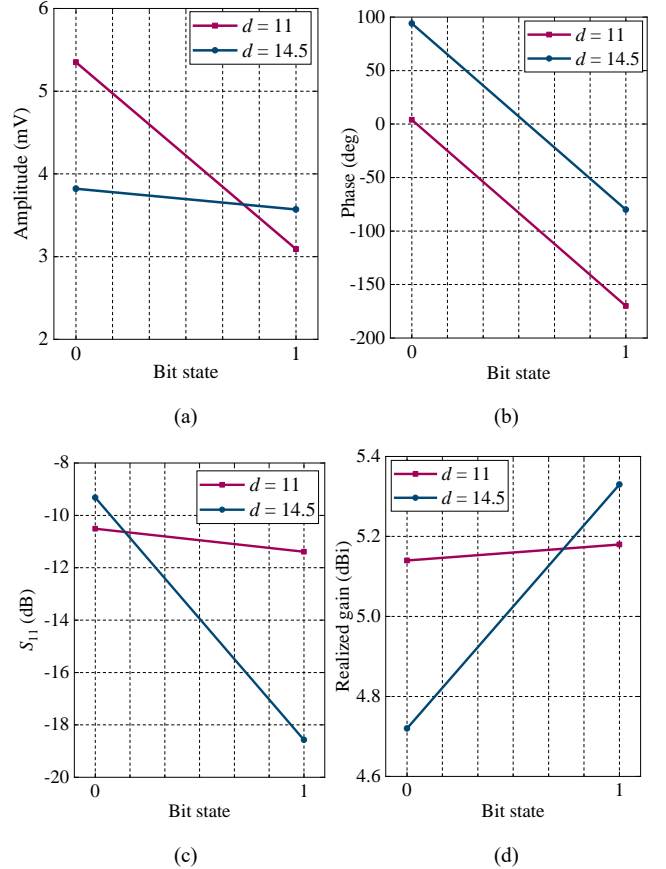

Fig. 9. Simulated radiation and scattering performance of elements I and II when the PIN diode is on and off: (a) scattering field amplitude, (b) scattering field phase, (c) antenna reflection coefficient, (d) antenna gain.

Then, a new 1×16 one-dimensional antenna array is randomly constructed using element I and II, as shown in Fig. 10. By employing equations (14) and (15), the required 1-bit phase distributions for different scattering beam scanning angles are calculated. Fig. 11(a) illustrates the simulated single-beam scanning performance of the antenna's scattering. It is observed that after adding the pre-phase, a good single-beam scanning performance up to $\theta_0 = 45^o$ for antenna array's in-band co-polarization scattering is achieved. Simultaneously, the radiation proficiency of the antenna array demonstrates substantial constancy, presenting a gain of approximately 16.3 dBi, as shown in Fig. 11(b). Specifically, the simulated scattering patterns at $\theta_0 = 30^o$ and $\theta_0 = 45^o$ of the designed antenna array with and without pre-phase are compared in Fig. 12. The results affirm the elimination of symmetric dual scattering beams post the addition of pre-phase, culminating in the attainment of the anticipated single-beam scanning performance.

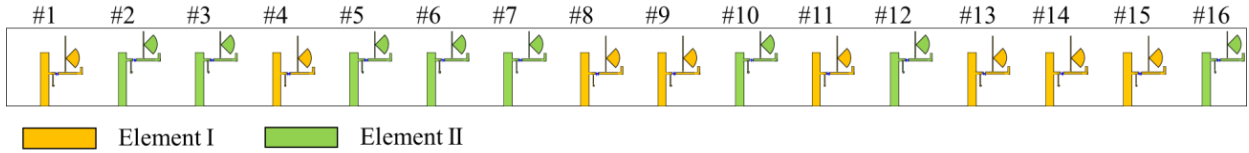

Fig. 10. 1×16 one-dimensional antenna array using element I and II.

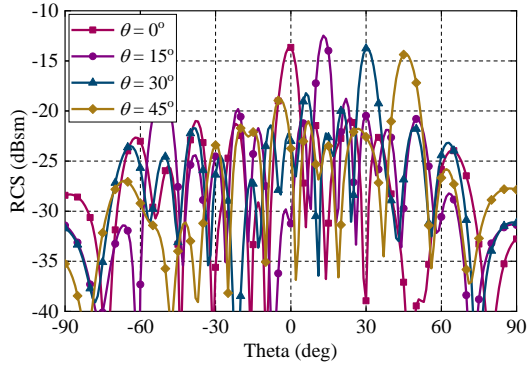

(a)

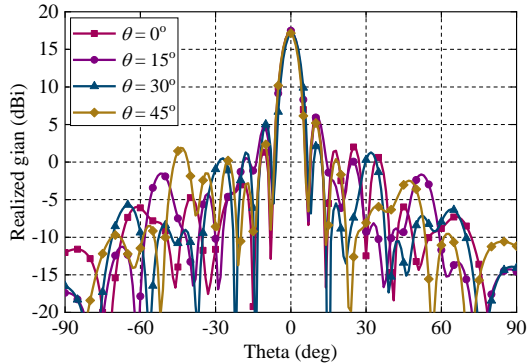

(b)

Fig. 11. Simulated scattering and radiation performance of the designed antenna array of different phase distributions, (a) scattering performance, (b) radiation performance

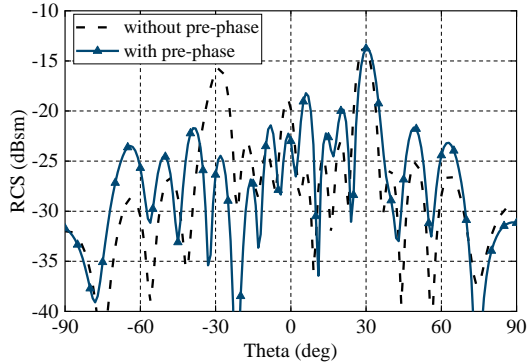

(a)

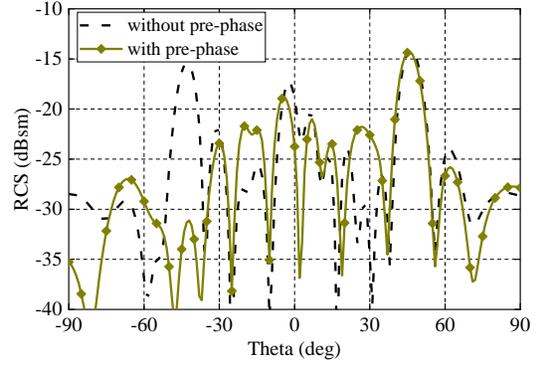

(b)

Fig. 12. Comparison of simulated dual-beam and single-beam scanning, (a) $\theta_0 = 30^o$, (b) $\theta_0 = 45^o$.

## V. FABRICATION AND MEASUREMENT

A prototype of the designed 1×16 antenna array is fabricated and assembled to validate the 1-bit reconfigurable single-beam scanning of the in-band co-polarization scattering, as shown in Fig. 13. The prototype has an aperture of 410 mm×30 mm and integrates 16 PIN diodes (SMP1340-040LF). Additionally, there are 16 pin connectors located on the side of the array, which are connected to the FPGA control board to enable electronic control of each element's state.

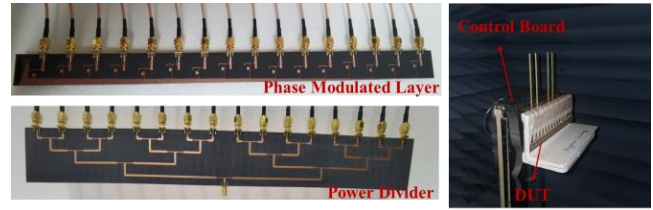

Fig. 13. Measurement setup for radiation performance.

### A  Radiation Performance

For the radiation performance measurement, a 1-to-16 power divider is used to provide an equal amplitude and phase feed. The port of this power divider is connected to the antenna array port through a coaxial line. The radiation performance is measured in an anechoic chamber, to obtain the reflection coefficients and gains under different phase codes for scattering beam scanning, as depicted in Fig. 13.

The Measured radiation performance of the designed antenna array under different scattering beam scanning codes are illustrated in Fig. 14. Compared with the simulation results, the measured operating frequency of the antenna is slightly offset to the low frequency, which is basically caused by fabricated errors. In addition, the measured gain is about 14.2 dBi, which is about 2 dB lower than the simulated gain, where the loss caused by the coaxial line is about 1.5 dB. In general, the radiation performance of antenna is relatively stable when regulating the in-band co-polarization scattering of antenna, which is in line with the design objectives.

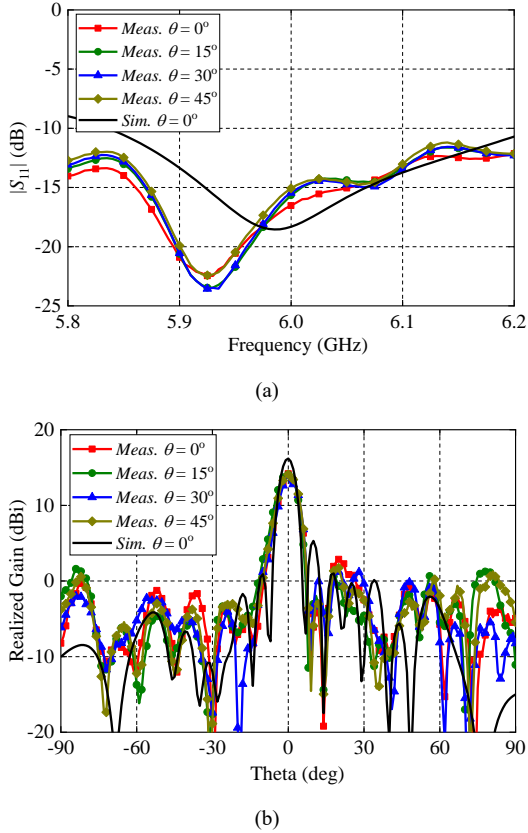

Fig. 14. Measured radiation performance of the designed antenna array under different scattering beam scanning codes. (a) reflection coefficients, (b) realized gain.

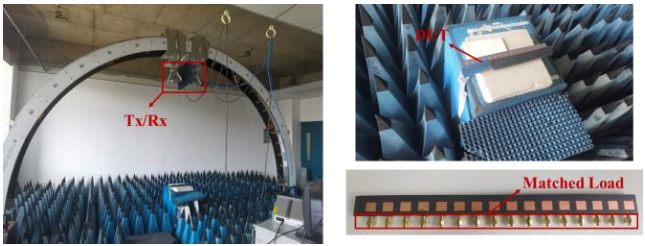

Fig. 15. Measurement setup for scattering performance

### B  Scattering Performance

To assess the scattering performance, the feed ports of the antenna array are terminated with a 50 Ω matching load. An FPGA control board is used to control the bias voltage of the PIN diode on each element to achieve different phase coding. A transmitting antenna, denoted as Tx, serves to normally irradiate the designed antenna array, while the receiving antenna, Rx, rotates around to measure the scattering pattern under different phase codes, verifying the single-beam scanning performance of the scattering, as depicted in Fig. 15.

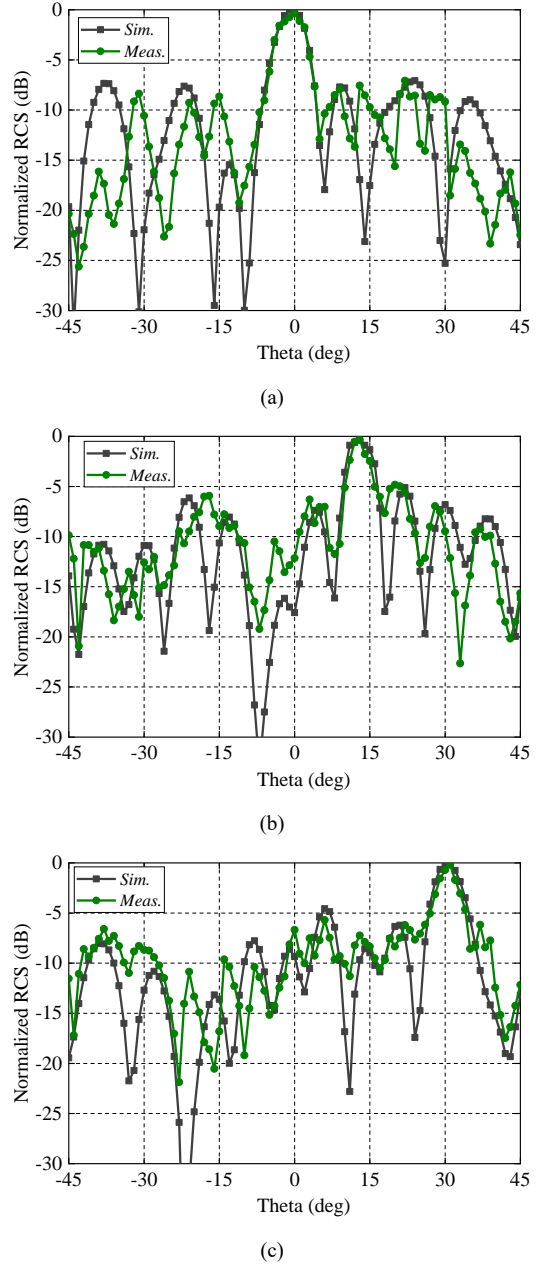

Fig. 16. Measured scattering performance of the designed antenna array. Single beam at (a) $\theta_0 = 0°$, (b) $\theta_0 = 15°$, (b) $\theta_0 = 30°$.

It should be mentioned that due to the limitations of the arch frame, the range of the scattering pattern is confined within $\pm 45°$, and the measured beam scanning angles include $0°$, $15°$ and $30°$. In addition, this method is difficult to measure the exact value of RCS, so it is normalized when comparing with the simulation results, as illustrated in Fig. 16. The measured results show that the designed antenna array can achieve accurate single beam scanning of the in-band co-polarization scattering. Thus, the RCS in the expected direction is enhanced while the backward RCS is effectively reduced, which reduces the probability of adversary detection and facilitating the coordination of ally equipment.

## VI. CONCLUSION

This paper demonstrates a novel approach to modulating the in-band co-polarization scattering of an antenna array through alterations in load impedance. By incorporating a PIN diode into the load impedance, 1-bit reconfiguration of the scattering field is achieved, and the single beam scanning performance of the scattering is realized by adding a pre-phase. A 1×16 linear antenna array is designed, fabricated and measured to validate the proposed methodology. The results indicate successful attainment of single scattering beam scanning along with stable radiation performance, under the condition of same frequency band and same polarization. Consequently, this method offers a viable solution for RCS regulation in antennas located on low observable platforms, enabling enhanced stealth capabilities by reducing the probability of detection from adversaries and facilitating harmonization among allied apparatuses.

## REFERENCES

[1]. A. K. Bhattacharyya and D. L. Sengupta, *Radar Cross Section Analysis and Control*. Norwood, MA, USA: Artech House, 1991.

[2]. Q. Lv, C. Jin, B. Zhang, P. Zhang, J. Wang, N. Kang, and B. Tian, "Wideband dual-polarized microwave absorber at extremely oblique incidence," *IEEE Trans. Antennas Propag.*, vol. 71, no. 3, pp. 2497-2506, Mar. 2023.

[3]. P. Gu, Z. Cao, Z. He, and D. Ding, "Design of ultrawideband RCS reduction metasurface using space mapping and phase cancellation," *IEEE Antennas Wireless Propag. Lett.*, vol. 22, pp. 1386-1390, Jun. 2023.

[4]. R. Zaker and A. Sadeghzadeh, "A low-profile design of polarization rotation reflective surface for wideband RCS reduction," *IEEE Antennas Wireless Propag. Lett.*, vol. 18, pp. 1794-1798, Sep. 2019.

[5]. B. Zhang, C. Jin and Z. Shen, "Low-profile broadband absorber based on multimode resistor-embedded metallic strips," *IEEE Trans. Microw. Theory Tech.*, vol. 68, no. 3, pp. 835-843, Mar. 2020.

[6]. Y. Li, P. Gu, Z. He, Z. Cao, J. Cao, K. Leung, and D. Ding, "An ultra-wideband multilayer absorber using an equivalent circuit-based approach," *IEEE Trans. Antennas Propag.*, vol. 70, no. 12, pp. 11911-11921, Dec. 2022.

[7]. Z. Ma, C. Jiang, W. Cao, J. Li, and X. Huang, "An ultrawideband and high-absorption circuit-analog absorber with incident angle-insensitive performance," *IEEE Trans. Antennas Propag.*, vol. 70, no. 10, pp. 9376-9384, Oct. 2022.

[8]. Y. Cheng, J. Feng, C. Liao, and X. Ding, "Analysis and design of wideband low-RCS wide-scan phased array with AMC ground," *IEEE Antennas Wireless Propag. Lett.*, vol. 20, pp. 209-213, Feb. 2021.

[9]. J. Xue, W. Jiang, and S. Gong, "Chessboard AMC surface based on quasi-fractal structure for wideband RCS reduction," *IEEE Antennas Wireless Propag. Lett.*, vol. 17, pp. 201-204, Feb. 2018.

[10]. D. Sang, Q. Chen, L. Ding, M. Guo, and Y. Fu, "Design of checkerboard AMC structure for wideband RCS reduction," *IEEE Trans. Antennas Propag.*, vol. 67, no. 4, pp. 2604-2612, Apr. 2019.

[11]. B. Zhang, C. Jin, and Z. Shen, "Absorptive frequency-selective reflector based on bent metallic strip embedded with chip-resistor," *IEEE Trans. Antennas Propag.*, vol. 68, no. 7, pp. 5736-5741, Jul. 2020.

[12]. B. Zhang, C. Jin, Q. Lv, J. Chen, and Y. Tang, "Low-RCS and wideband reflectarray antenna with high radiation efficiency," *IEEE Trans. Antennas Propag.*, vol. 69, no. 7, pp. 4212-4216, Jul. 2021.

[13]. Q. Lv, C. Jin, B. Zhang, and Z. Shen, "Hybrid absorptive-diffusive frequency selective radome," *IEEE Trans. Antennas Propag.*, vol. 69, no. 6, pp. 3312–3321, Jun. 2021.

[14]. X. Wang, P. Qin, and R. Jin, "Low RCS transmitarray employing phase controllable absorptive frequency-selective transmission elements," *IEEE Trans. Antennas Propag.*, vol. 69, no. 4, pp. 2398-2403, Apr. 2021.

[15]. H. Huang, C. Hua, and Z. Shen, "Absorptive frequency-selective transmission structures based on hybrid FSS and absorber," *IEEE Trans. Antennas Propag.*, vol. 70, no. 7, pp. 5606-5613, Jul. 2022.

[16]. L. Gan, W. Jiang, Q. Chen, X. Li, and Z. Zhou, "Analysis and reduction on in-band RCS of Fabry-Perot antennas," *IEEE Access*, vol. 8, pp. 146697-146706, 2020.

[17]. Y. Zhao, J. Gao, X. Cao, T. Liu, L. Xu, X. Liu, and L. Cong, "In-band RCS reduction of waveguide slot array using metasurface bars," *IEEE Trans. Antennas Propag.*, vol. 65, no. 2, pp. 943-947, Feb. 2017.

[18]. W. Xu, J. Wang, M. Chen, Z. Zhang, and Z. Li, "A novel microstrip antenna with composite patch structure for reduction of in-band RCS," *IEEE Antennas Wireless Propag. Lett.*, vol. 14, pp. 139-142, 2015.

[19]. P. Yang, F. Yan, F. Yang, and T. Dong, "Microstrip phased-array in-band RCS reduction with a random element rotation technique," *IEEE Trans. Antennas Propag.*, vol. 64, no. 6, pp. 2513-2518, Jun. 2016.

[20]. K. Ji, X. Cao, J. Gao, H. Yang, T. Li, L. Jidi, Z. Zhang, and Jia Lu, "Design of low profile ATFSS and antenna with in-band and out-of-band RCS reduction," *IEEE Trans. Antennas Propag.*, vol. 70, no. 12, pp. 11537-11547, Dec. 2022.

[21]. L. Yin, P. Yang, Y. Gan, F. Yang, S. Yang, and Z. Nie, "A low cost, low in-band RCS microstrip phased-array antenna with integrated 2-bit phase shifter," *IEEE Trans. Antennas Propag.*, vol. 69, no. 8, pp. 4517-4526, Aug. 2021.

[22]. P. Wang, Y. Jia, W. Hu, Y. Liu, H. Lei, H. Sun, and T. Cui, "Circularly polarized polarization conversion metasurface-inspired antenna array with low RCS over a wide band," *IEEE Trans. Antennas Propag.*, vol. 71, no. 7, pp. 5626-5636, Jul. 2023.

[23]. Y. Liu, Y. Jia, W. Zhang, Y. Wang, S. Gong, and G. Liao, "An integrated radiation and scattering performance design method of low-RCS patch antenna array with different antenna elements," *IEEE Trans. Antennas Propag.*, vol. 67, no. 9, pp. 6199-6204, Sep. 2019.

[24]. S. Xiao, S. Yang, Y. Chen, S. Qu, L. Sun, and J. Hu, "In-band scattering reduction of wideband phased antenna arrays with enhanced coupling based on phase-only optimization techniques," *IEEE Trans. Antennas Propag.*, vol. 68, no. 7, pp. 5297-5307, Jul. 2020.

[25]. Z. Zhang, F. Yang, Y. Chen, S. Qu, J. Hu, and S. Yang, "In-band scattering and radiation tradeoff of broadband phased arrays based on scattering-matrix approach," *IEEE Trans. Antennas Propag.*, vol. 69, no. 11, pp. 7486-7496, Nov. 2021.

[26]. P. Li, S. Qu, S. Yang, and J. Hu, "In-band SCS reduction of microstrip phased array based on impedance matching network," *IEEE Trans. Antennas Propag.*, vol. 70, no. 1, pp. 330-340, Jan. 2022.

[27]. H. Yang, T. Li, L. Xu, X. Cao, L. Jidi, Z. Guo, P. Li, and J. Gao, "Low in-band-RCS antennas based on anisotropic metasurface using a novel integration method," *IEEE Trans. Antennas Propag.*, vol. 69, no. 3, pp. 1239-1248, Mar. 2021.

[28]. Y. Lv, R. Wang, B. Wang, and Z. Chen, "Anisotropic complementary metantenna for low sidelobe radiation and low in-band co-polarized scattering using characteristic mode analysis," *IEEE Trans. Antennas Propag.*, vol. 70, no. 11, pp. 10177-10186, Nov. 2022.

[29]. J. Zhang, Y. Liu, Y. Jia, and R. Zhang, "High-gain Fabry–Perot antenna with reconfigurable scattering patterns based on varactor diodes," *IEEE Trans. Antennas Propag.*, vol. 70, no. 2, pp. 922-930, Feb. 2022.

[30]. Y. Liu, W. Zhang, Y. Jia, and A. Wu, "Low RCS antenna array with reconfigurable scattering patterns based on digital antenna units," *IEEE Trans. Antennas Propag.*, vol. 69, no. 1, pp. 572–577, Jan. 2021.

[31]. Z. Zhang, F. Yang, Y. Chen, S. Qu, and J. Hu, "Low-scattering phased arrays with reconfigurable scattering patterns based on independent control of radiation and scattering," *IEEE Trans. Antennas Propag.*, vol. 71, no. 6, pp. 5057-5066, Jun. 2023.

[32]. R. B. Green, *The General Theory of Antenna Scattering*, Antenna Laboratory. Columbus, OH, USA: Ohio State Univ., Nov. 1963.

[33]. M. Wang, S. Xu, F. Yang, and M Li, "A 1-bit bidirectional reconfigurable transmit-reflect-array using a single-layer slot element with PIN diodes," *IEEE Trans. Antennas Propag.*, vol. 6, no. 9, pp. 6205-6210, Feb. 2021.

[34]. J. Yin, Q. Wu, Q. Lou, H. Wang, Z. Chen, and W. Hong, "Single-beam 1 bit reflective metasurface using prephased unit cells for normally incident plane waves," *IEEE Trans. Antennas Propag.*, vol. 68, no. 7, pp. 5496-5504, July 2020.